\documentclass[11pt,a4paper]{article}
\pdfoutput=1
\usepackage{a4wide}
\usepackage{amsfonts}
\usepackage{amssymb}
\usepackage{amsmath}
\usepackage{graphicx}
\usepackage{booktabs}
\usepackage{array}
\usepackage{color}
\usepackage{ulem}
\usepackage[small,bf]{caption}
\usepackage{cite}
\usepackage[usenames,dvipsnames]{xcolor}
\usepackage{subfigure}
\usepackage{verbatim}
\usepackage{dsfont}
\usepackage{bbm}
\usepackage{hyperref}
\def\eps{\epsilon}
\def\1{\mathbf{1}}
\def\3{\mathbf{3}}
\def\2{\mathbf{2}}

\def\ii{\mathrm{i}}
\def\MSbar{\overline{\text{MS}}}
\def\DRbar{\overline{\text{DR}}}

\allowdisplaybreaks[1]

\numberwithin{equation}{section}

\newcounter{mysubequation}[equation]

\definecolor{pink}{rgb}{1.,.2,.8}
\definecolor{cinnamon}{rgb}{0.82, 0.41, 0.12}

\begin{document}
\begin{titlepage}

\vspace*{-15mm}
\begin{flushright}
TTP18-038 \\
NCTS-PH-1816
\end{flushright}
\vspace*{0.7cm}

\begin{center}
  { \bf\LARGE Confronting SUSY SO(10) with \\[0.2em] updated Lattice and Neutrino Data} 
\\[8mm]
Thomas Deppisch$^{\, a}$ \footnote{E-mail: \texttt{thomas.deppisch@kit.edu}},
Stefan Schacht$^{\, b, c}$ \footnote{E-mail: \texttt{ss3843@cornell.edu}},
Martin Spinrath$^{\, d, e}$ \footnote{E-mail: \texttt{spinrath@phys.nthu.edu.tw}}
\\[1mm]
\end{center}
\vspace*{0.50cm}
\centerline{$^{a}$ \it Institut f\"ur Theoretische Teilchenphysik, Karlsruhe Institute of Technology,}
\centerline{\it Engesserstra\ss{}e 7, D-76131 Karlsruhe, Germany}
\vspace*{0.2cm}
\centerline{$^{b}$ \it Department of Physics, LEPP, Cornell University, Ithaca, NY 14853, USA}
\centerline{$^{c}$ \it Dipartimento di Fisica, Universit\`a di Torino \& INFN,}
\centerline{\it Sezione di Torino, I-10125 Torino, Italy}
\vspace*{0.2cm}
\centerline{$^{d}$ \it Department of Physics, National Tsing Hua University, Hsinchu 30013, Taiwan}
\centerline{$^{e}$ \it Physics Division, National Center for Theoretical Sciences, Hsinchu 30013, Taiwan}
\vspace*{1.20cm}

\begin{abstract}
\noindent
We present an updated fit of supersymmetric SO(10) models to quark and lepton masses and mixing parameters.
Including latest results from lattice QCD determinations of quark masses and neutrino oscillation data,
we show that fits neglecting supersymmetric threshold corrections are strongly disfavoured in our setup.
Only when we include
these corrections we find good fit points. We present $\chi^2$-profiles for the threshold parameters, 
which show that in our setup the thresholds related to the third generation of fermions
exhibit two rather narrow minima.
\end{abstract}

\end{titlepage}

\setcounter{footnote}{0}

\section{Introduction}
\label{sec:Introduction}

Since the advent of Maxwell's theory of electromagnetism it is a common dream in physics
to unify all interactions into one single theory. This dream is particularly persistent in particle physics
where Grand Unified Theories (GUTs) -- which unify three of the fundamental forces of nature --  have
been a major guiding principle in the last couple of decades. To be precise in this paper we will
focus on GUTs based on the SO(10) gauge group first proposed in the 1970s \cite{Georgi:1974my, Fritzsch:1974nn}. 
This choice contains all the essential features interpreted
as hints towards unification, like charge quantization, anomaly cancellation or smallness of neutrino masses.

SO(10) GUTs work especially well in the context of low energy supersymmetry (SUSY) 
which helps significantly to unify the gauge couplings, see, e.g., \cite{Amaldi:1991cn}. In fact,
supersymmetry is part of the dream of unification as it unifies the concepts of bosons and fermions
or the concepts of matter and forces. It also provides an elegant mechanism to stabilize mass hierarchies and provides a dark matter candidate on top.

In this paper though we will not discuss all features and aspects of SUSY SO(10) models in general.
We focus on the Yukawa sector, which is interesting because this sector is constrained by low
energy observables, namely the observed fermion masses and mixing parameters. There has been
some tremendous progress in the last couple of years in particular for light quark masses from lattice
computations and neutrino masses and mixing from oscillation experiments. We will use this updated
information to provide an updated fit to a minimal SUSY SO(10) Yukawa sector.

There is a long history of fitting SO(10) to fermion masses, see, e.g., \cite{
Aulakh:1982sw, Clark:1982ai,
Ananthanarayan:1991xp,
9209215,
9306297,
9405399,
9710371,
9906433,
0004031,
0110310,
0205066,
0206239,
0210207,
0303055,
0305166,
0306242,
0308197,
0402122,
0402140,
0405074,
0406117,
0406262,
0504241,
0505200,
0511352,
0512224,
0605006,
0607197,
0612021,
0702284,
0704.1248,
0710.3945,
0710.4018,
0807.0917,
0809.1069,
1012.2697,
1102.5148,
1107.2378,
1202.4012,
1203.0829,
1203.1403,
1305.1001,
1306.4468,
Babu:2016bmy, Babu:2018tfi}
Some works focussed on particular SO(10) models,
while others focussed on a general SO(10) Yukawa sector only.
These efforts have always been a difficult task due to the large number of parameters and observables.
For that reason many groups had to resort to some simplifications or estimates. For instance, the recent
fits by Dueck and Rodejohann \cite{1306.4468} neglect in the SUSY case threshold corrections
which are known to give sizeable, important corrections, e.g., Refs.~\cite{SUSYthresholds, 0704.1248, 1102.5148, 0702267, 0804.0717}. One of the main
improvements of our work compared to previous studies is to include these threshold corrections
in terms of three parameters and to provide $\chi^2$-profiles for them for the first time, to our knowledge.

Our paper is organised as follows: In Section~\ref{sec:Fitting} we describe in detail how we match the
SO(10) Yukawa couplings at the high scale to the fermion masses and mixing parameters at low energies
including the fit procedure.
In Section~\ref{sec:Results} we discuss our results. First we present the global minima of our fits and 
the pulls to identify which observables are potentially driving some tensions of the data with our
setup. Then we present the $\chi^2$-profiles for the SUSY threshold corrections before we summarise
and conclude in Section~\ref{sec:Summary}.  
We provide additional information in two appendices. 
Appendix~\ref{sec:observables} gives additional numerical values for Standard Model (SM) observables, 
and Appendix~\ref{sec:gutscaleparameters} the values of the GUT scale fit parameters at the global
and local minima.

\section{Fitting GUT Scale Yukawa Couplings to Fermion Masses}
\label{sec:Fitting}

Besides the unification of the gauge sector, SO(10) GUTs also predict the 
unification of the Yukawa sector by arranging all matter fields in the 
spinorial $\mathbf{16}$ representation. Of course, these constraints hold
only at the GUT scale and effects of GUT breaking and of the renormalisation
group equations (RGEs) will alter the
relations among the Yukawa couplings.

Below the GUT scale, important physics happens around the seesaw scale where
the heavy right-handed neutrinos are integrated out one after another.
Moreover, SUSY breaking effects have to be taken into account in
supersymmetric GUTs. Especially
for large values of $\tan \beta$, threshold corrections from SUSY particles
have sizeable effects \cite{SUSYthresholds}. In our analysis, we choose
the matching scale between the SM and its minimal supersymmetric
extension (MSSM) to be $M_{\text{SUSY}} = 1$~TeV. At this scale we define as well
the $\chi^2$-function to fit the GUT parameters to the SM Yukawa
couplings. Therefore we first need to evolve
the fermion masses and mixing parameters to $M_{\text{SUSY}}$
using the SM $\beta$-functions. We describe the procedure in greater detail in the following.

\subsection{Yukawa Couplings in SO(10) GUTs}
\label{sec:yukawacouplingsSO10}

In this work we restrict ourselves to renormalisable SO(10) GUTs.
This has a stronger predictive power than SU(5) or non-renormalisable
GUTs. 
Furthermore we assume the standard embedding of the matter fields
into the spinorial $\mathbf{16}$ of SO(10). This restricts
the Higgs representations relevant for fermion masses to a $\mathbf{10}$,
$\mathbf{120}$ and/or $\overline{\mathbf{126}}$ of SO(10).

In SUSY SO(10), the most general renormalisable superpotential
describing Yukawa interactions is then given by
\begin{equation}
  \mathcal{W} = Y^{ij}_{10}\ \mathbf{16}_i \cdot\mathbf{16}_j \cdot \mathbf{10}_H
  + Y^{ij}_{120}\ \mathbf{16}_i \cdot \mathbf{16}_j \cdot \mathbf{120}_H
  + Y^{ij}_{126}\ \mathbf{16}_i \cdot \mathbf{16}_j \cdot \overline{\mathbf{126}}_H \;,
\end{equation}
where the $\mathbf{16}_i$ are the three generations of matter fields
with flavour indices $i,j=1,2,3$. The fields $\mathbf{10}_H$, $\mathbf{120}_H$
and $\overline{\mathbf{126}}_H$ are GUT representations containing Higgs fields
and we assume maximum one of each.
Due to the SO(10) gauge structure the Yukawa matrices $Y^{ij}_{10}$ and $Y^{ij}_{126}$
are symmetric while $Y^{ij}_{120}$ is antisymmetric in the flavour indices $i$
and $j$. Without loss of generality we apply an unphysical flavour rotation
to choose $Y_{10}$ real and diagonal.

After breaking SO(10) at the high scale $M_{\text{GUT}} = 1.353\times 10^{16}$~GeV,
given by gauge coupling unification\footnote{For details see
  Appendix \ref{sec:gutscaleparameters}.},
the Yukawa matrices are matched to the MSSM Yukawa matrices $Y_x$, $x = u, d, e, \nu$,
the Majorana mass matrix for the right handed neutrinos $M$ and the Wilson
coefficient of the Weinberg operator $\kappa$, see for example Ref.~\cite{1306.4468}
\begin{align}
  Y_u &= r\,(Y_{10}+s\, Y_{126} + \ii \, t_u\, Y_{120}) \;, \label{eq:Yu_GUT}\\
  Y_d &= Y_{10}+ Y_{126} + \ii\, t_d Y_{120} \;, \label{eq:Yd_GUT}\\
  Y_\nu &= r\,(Y_{10}-3s\, Y_{126} + \ii \, t_\nu\, Y_{120}) \;, \label{eq:Yv_GUT}\\
  Y_e &= Y_{10}-3\, Y_{126} + \ii \, t_e\, Y_{120} \;, \label{eq:Ye_GUT}\\
  M &=  r_R\, Y_{126} \;, \label{eq:MN_GUT}\\
  \kappa &=  r_L\, Y_{126} \label{eq:Ka_GUT} \;.
\end{align}
The parameters $s$, $t_u$, $t_d$, $t_\nu$, $t_e$ correspond to the mixing of
GUT scale Higgs doublets into the MSSM Higgs doublets $H_u$, $H_d$ whereas $r$,
$r_R$, $r_L$ correspond to the vacuum expectation values (vevs) of the GUT scale Higgs fields and can be
chosen real without loss of generality.

A minimal realistic choice omits the $\mathbf{120}$ representation (i.e.,
the parameters $t_u$, $t_d$, $t_\nu$, $t_e$) and $r_L$.
Previous fits to
fermion observables have shown that the $\mathbf{120}$ does not significantly
improve the fit to fermion masses in the SM, e.g., Ref.~\cite{1306.4468},
so we will not include it here ($Y_{120} = 0$). Furthermore, in Ref.~\cite{0605006} it was
argued that while a generic fit to fermion masses prefers a mix of type-I seesaw (parametrised by $r_R$)
and type-II seesaw (parametrised by $r_L$), a fit which includes the GUT potential
favours scenarios where type-I is dominant over type-II. Hence, we will assume $r_L =0$.
Thus we end up with a set of 19 GUT scale parameters: $Y_{10}$ (three parameters),
 $Y_{126}$ (twelve parameters), $r_R$ (one parameter), $r$ (one parameter) and $s$ (two parameters).

Before we discuss our numerical implementation and the results we want
to give a few comments on Yukawa couplings from higher-dimensional operators.
In this study we explicitly assume that no higher-dimensional operators
correct the Yukawa coupling relations beyond the ones present at
renormalisable level. Especially, in the context of supergravity one might
expect higher-dimensional operators to be present in the superpotential
suppressed by powers of $M_{\text{GUT}}/{M_{\text{Planck}}} =
\mathcal{O}(0.01)$. Such higher-dimensional operators could alter
the Yukawa relations if they have a non-trivial gauge structure.

It is an open question if quantum gravity introduces GUT non-singlets
at the Planck scale which couple to ordinary matter. 
The presence of non-singlets would imply that quantum
gravity knows something about flavour. If it would be flavour-blind all Yukawa couplings
should be at least of order $M_{\text{GUT}}/{M_{\text{Planck}}} =
\mathcal{O}(0.01)$. However, this is not the case unless there are some large
cancellations at work. Hence, we interpret
the smallness of the electron mass as an indication that these
higher-dimensional operators are negligibly small and we do not
consider them here. Nevertheless, in a supersymmetric GUT theory of flavour
they can be the dominant operators and very sensible, for a recent review, see, e.g.,
Ref.~\cite{King:2017guk}.

\subsection{RGE Running from the GUT Scale to the SUSY Scale}

The values of the Yukawa couplings at different energy scales are related by the
renormalisation group equations via
\begin{equation}
  Y(\mu_2) = Y(\mu_1) + \int\limits_{\log\mu_1}^{\log\mu_2} \text{d}\log{\mu}\, \beta_Y(\mu),
\end{equation}
with the beta function being defined as $\beta_Y(\mu) = \text{d}Y / \text{d}\log\mu$.
At one- and two-loop level they are well-known for both the SM and the MSSM with additional
right-handed neutrinos, see, e.g., Ref.~\cite{Antusch:2002ek}. We also include the
running of the Weinberg operator at the one-loop level.

For every point in the GUT parameter space the RGEs have to be solved, i.e.\
numerically integrated. For a correct treatment of the seesaw mechanism, we consecutively
integrate out the right-handed neutrinos as described in Ref.~\cite{Antusch:2005gp} and
implemented in the \texttt{Mathematica} package \texttt{REAP} presented therein.
\texttt{REAP} provides a useful tool for cross checking our calculations.
We implement the numerical procedure in \texttt{C++}
using the template libraries \texttt{odeint} \cite{Ahnert:2011} and \texttt{eigen3} \cite{eigenweb}.
This speeds up the calculation of the RGE running by up to a factor of 100 compared to
the \texttt{Mathematica} version of \texttt{REAP} on our test desktop machine (Intel i5-4590 CPU, 3.30 GHz).

\subsection{Fermion masses at the SUSY scale}
\label{sec:fermionmasses}
 
\begin{table}
  \begin{center}
    \caption{
      Input data for SM parameters in the $\MSbar$ scheme from lattice QCD determinations with $N_f=2+1+1$ taken from Ref.~\cite{Aoki:2016frl}. }
    \label{tab:inputdata}
\begin{tabular}{cc}
\toprule
Input    	     & Value   \\
\midrule
$m_u (\text{2 GeV})$  & 2.36(24)  \, \text{MeV}\\
$m_d (\text{2 GeV})$  & 5.03(26)  \, \text{MeV}\\
$m_s(\text{2 GeV})$   & 93.9(1.1) \, \text{MeV}  \\
$m_c(\text{3 GeV})$   & 996(25)   \, \text{MeV}  \\
$\bar m_b(\bar m_b)$  & 4.190(21) \, \text{GeV}  \\
$\alpha_s^{(5)}(M_Z)$ & 0.1182(12) \\
\bottomrule
\end{tabular}
\end{center}
\end{table}

Our low scale input for the quark masses and $\alpha_s$ is given in Table~\ref{tab:inputdata}.
We perform the QCD RGE evolution to $M_Z= 91.1876\, \text{GeV}$ \cite{Olive:2016xmw} and the 
matching from 4 to 5 as well as from 5 to 6 active flavours with the \texttt{Mathematica} package \texttt{RunDec} 
\cite{Chetyrkin:2000yt} to four-loop accuracy. Our results are shown in Table~\ref{tab:RGEresult6}.
Errors from variation of the bottom and top quark scales are negligible
compared to the experimental uncertainties. The Yukawa couplings at $M_Z$ can be obtained
from $y=m/v$, with $v=174.104\, \text{GeV}$. As additional input at $M_Z$, we use the values
for the gauge couplings $g_1$ and $g_2$ as well as the values for the lepton Yukawa couplings
from Ref.~\cite{Antusch:2013jca}. Regarding the input for the CKM matrix, we use the ICHEP2016
update from CKMfitter~\cite{Charles:2004jd}. Corresponding results from UTfit are in good
agreement~\cite{Bona:2005vz}.

\begin{table}
  \begin{center}
    \caption{
      Results of the QCD RG evolution for six active flavours and 
      a top-quark pole mass $m_{t,\text{pole}}=174.2 \pm 1.4\, \text{GeV}$ \cite{Olive:2016xmw}.}
    \label{tab:RGEresult6}
\begin{tabular}{cc}
\toprule
Result at $M_Z$ & Value   \\
\midrule
  $m_u^{(6)}(M_Z)$ 	 & 1.36(15)   \,MeV \\
  $m_d^{(6)}(M_Z)$ 	 & 2.90(11)   \,MeV \\
  $m_s^{(6)}(M_Z)$ 	 & 54.05(63)  \,MeV \\
  $m_c^{(6)}(M_Z)$ 	 & 635(16)    \,MeV \\
  $m_b^{(6)}(M_Z)$ 	 & 2.866(14)  \,GeV \\
  $m_t^{(6)}(M_Z)$ 	 & 172.3(1.5) \,GeV \\
  $\alpha_s^{(6)}(M_Z)$	 & 0.1170(12)       \\
  \bottomrule
\end{tabular}
\end{center}
\end{table}

In order to perform the RGE evolution from $M_Z$ to $M_{\mathrm{SUSY}}$, we use the
two-loop RGEs for the SM Yukawas and gauge couplings \cite{Machacek:1983fi}. For solving
the RGEs at two-loop also the Higgs quartic coupling is needed which we get from the PDG
average of the Higgs mass measurement $m_{\text{Higgs}} = 125.09 \pm 0.24\,\text{GeV}$~\cite{Olive:2016xmw}.  
For the off-diagonal elements of the Yukawa matrices we use the standard parametrisation
of the PDG. Further details can be found in Appendix~\ref{sec:observables}.

We treat the uncertainties of the experimental data as Gaussian and
symmetrise them arithmetically where necessary. For the propagation of uncertainties we sample normally
distributed random numbers at $M_Z$. Each sample point is then evolved to the respective energy scale.
There, we fit a normal distribution to the set of sample points. Thus, also non-linear effects, especially the
influence of the top quark Yukawa coupling, and interdependencies in the RGE running are taken into account.
This has the effect that the relative errors
are larger at higher energy scales due to the error of the top quark mass measurement.

For our global fit we set $M_{\mathrm{SUSY}}=1$ TeV. However, for reference,
in Table~\ref{tab:SM} we additionally
list also the SM Yukawa and gauge couplings in the $\MSbar$ scheme at $M_Z$, 3 TeV and 10 TeV  for reference.
The same results converted to the $\DRbar$ scheme can be found in Table~\ref{tab:DRbar}.

\begin{table}
  \centering
  \caption{\textbf{SM observables in the $\MSbar$ scheme at $\boldsymbol{M_Z}$, 1\, TeV, 3\, TeV and 10\, TeV.}
    Quark masses and $\alpha_s$ are taken from lattice determinations~\cite{Aoki:2016frl} and evolved
    to $M_Z$ with \texttt{RunDec} \cite{Chetyrkin:2000yt}. All quark Yukawa couplings and $\alpha_s$
    correspond to six active flavours.
    Lepton Yukawas and electroweak gauge couplings are taken from Ref.~\cite{Antusch:2013jca}, CKM
    parameters from the ICHEP 2016 update of Ref.~\cite{Charles:2004jd}.
    Note that we use the GUT normalisation for $g_1$.
    The running between $M_Z$ and 1, 3 and 10 TeV is
    performed using the SM RGEs at two-loop. Gaussian errors are propagated applying Bayes' theorem.
    }
  \small
  \begin{tabular}{ccccc}\toprule
     & $M_Z$ & 1 TeV & 3 TeV & 10 TeV \\\midrule
    $g_1$ & $0.461425^{+0.000044}_{-0000043}$ & $0.467773\pm 0.000045$& $0.470766\pm 0.000046$& $0.474109\pm 0.000047$\\ \addlinespace
    $g_2$ & $0.65184^{+0.00018}_{-0.00017}$ &  $0.63935\pm 0.00016$ & $0.63383\pm 0.00016$& $0.62792\pm 0.00016$ \\ \addlinespace
    $g_3$ & $1.2127\pm 0.0061$ &  $1.0549 \pm 0.0040$ & $1.0009\pm 0.0034$& $0.9503\pm 0.0029$\\\midrule
    $y_u/10^{-6}$ & $7.80\pm 0.86$ & $6.73\pm 0.74$ & $6.37\pm 0.70$& $6.03\pm 0.66$\\ \addlinespace
    $y_c/10^{-3}$ & $3.646\pm 0.091$ & $3.147\pm 0.79$& $2.976\pm 0.074$ & $2.816\pm 0.071$\\ \addlinespace
    $y_t$ & $0.9897\pm 0.0086$ & $0.8723\pm 0.0088 $& $0.8317\pm 0.0088$& $0.7934\pm 0.0089$\\\midrule
    $y_d/10^{-5}$ & $1.663\pm 0.064$ & $1.438\pm 0.056$ & $1.361\pm 0.053$& $1.289\pm 0.050$\\  \addlinespace
    $y_s/10^{-4}$ & $3.104\pm 0.036$ & $2.685\pm 0.032$& $2.541\pm 0.030$ & $2.407\pm 0.029$\\ \addlinespace
    $y_b/10^{-2}$ & $1.646\pm 0.0082$ & $1.3940\pm 0.0079$ & $1.3091\pm 0.0071$ & $1.2303\pm 0.0070$\\\midrule
    $y_e/10^{-6}$ & $2.794745^{+0.000015}_{-0.000016}$ & $2.8491\pm 0.0022$ & $2.8659\pm 0.0031$ & $2.8800\pm 0.0040$ \\ \addlinespace
    $y_\mu/10^{-4}$ & $5.899863^{+0.000019}_{-0.000018}$ & $6.0146\pm 0.0046$& $6.0501\pm 0.0065$& $6.080\pm 0.0086$\\ \addlinespace
    $y_\tau/10^{-2}$ & $1.002950^{+0.000090}_{-0.000091}$ & $1.02246\pm 0.00078$ & $1.0285\pm 0.0011$& $1.0336\pm 0.0014$ \\\midrule
    $\theta_{12}^{q}$ & $0.22704^{+0.00030}_{-0.00029}$ & $0.22704\pm 0.00029$ & $0.22704\pm 0.00029$& $0.22705\pm 0.00029$\\ \addlinespace
    $\theta_{13}^{q}/10^{-3}$ & $3.71^{+0.13}_{-0.14}$ & $3.79\pm 0.14$ & $3.82\pm 0.14$& $3.85\pm 0.14$\\ \addlinespace
    $\theta_{23}^{q}/10^{-2}$ & $4.181^{+0.047}_{-0.067}$ & $4.270\pm 0.058$ & $4.303\pm 0.058$& $4.337\pm 0.059$ \\ \addlinespace
    $\delta_{\text{CP}}^{q}$ & $1.143^{+0.011}_{-0.011}$ & $1.143\pm 0.011$ & $1.143\pm 0.011$& $1.143\pm 0.011$ \\\bottomrule
  \end{tabular}
  \label{tab:SM}
\end{table}

\begin{table}
  \centering
  \caption{\textbf{SM observables in the $\DRbar$ scheme at $\boldsymbol{M_Z}$, 1\, TeV, 3\, TeV and 10\, TeV.} 
  The results at 1\, TeV are included as input in our SO(10) fit. Quark masses and $\alpha_s$ are taken from
  lattice determinations~\cite{Aoki:2016frl} and evolved to $M_Z$ with \texttt{RunDec} \cite{Chetyrkin:2000yt}.
  All quark Yukawa couplings and $\alpha_s$ correspond to six active flavours. Lepton Yukawa couplings and electroweak gauge couplings
  are taken from Ref.~\cite{Antusch:2013jca}, CKM parameters from the ICHEP 2016 update of Ref.~\cite{Charles:2004jd}.
  Note that we use the GUT normalisation for $g_1$.
  The running
  between $M_Z$ and 1, 3 and 10 TeV is performed using the SM RGEs at two-loop. Gaussian errors are propagated
  applying Bayes' theorem. All gauge couplings and Yukawa couplings have been converted to
  $\DRbar$ at the given scale according to Ref.~\cite{Martin:1993yx}.
}
  \label{tab:DRbar}
  \small
  \begin{tabular}{ccccc}\toprule
     & $M_Z$ & 1 TeV & 3 TeV & 10 TeV \\ \midrule
    $g_1$ & $0.461425^{+0.000044}_{-0000043}$ & $0.467773\pm 0.000045$& $0.470766\pm 0.000046$& $0.474109\pm 0.000047$\\ \addlinespace
    $g_2$ & $0.65243\pm  0.00018$ & $0.63990\pm  0.00017$ & $0.63437\pm  0.00016$ & $0.62844\pm  0.00016$ \\ \addlinespace
    $g_3$ & $1.2185\pm  0.0062$ & $1.0587\pm  0.0040$ & $1.00415\pm  0.0034$ & $0.9530\pm  0.0029$  \\ \midrule
    $y_u/10^{-6}$ & $7.71\pm  0.85$ & $6.68\pm  0.74$ & $6.32\pm  0.70$ & $5.99\pm  0.66$  \\ \addlinespace
    $y_c/10^{-3}$ & $3.60\pm  0.090$ & $3.120\pm  0.078$ & $2.954\pm  0.074$ & $2.797\pm  0.071$  \\ \addlinespace
    $y_t$ & $0.9785\pm  0.0086$ & $0.8651\pm 0.0087$ & $0.8255\pm  0.0087$ & $0.7881\pm 0.0088$  \\ \midrule
    $y_d/10^{-5}$ & $1.64441\pm  0.064$ & $1.426\pm 0.056$ & $1.351\pm  0.053$ &  $1.281\pm  0.050$ \\ \addlinespace
    $y_s/10^{-4}$ & $3.06979\pm  0.036$ & $2.663\pm 0.032$ & $2.523\pm  0.030$ &  $2.392\pm  0.029$ \\ \addlinespace
    $y_b/10^{-2}$ & $1.6274\pm  0.0082$ & $1.3825\pm 0.0073$ & $1.2995\pm 0.0071$ & $1.2224\pm 0.0070$ \\ \midrule
    $y_e/10^{-6}$ & $2.796719\pm  0.000016$ & $2.8510\pm 0.0022$ & $2.8677\pm 0.0031$ &  $2.8818\pm  0.0041$ \\ \addlinespace
    $y_\mu/10^{-4}$ & $5.904029\pm  0.000019$ & $6.0186\pm 0.0046$ & $6.0540\pm  0.0065$ &  $6.0836\pm  0.0086$ \\ \addlinespace
    $y_\tau/10^{-2}$ & $1.003658\pm  0.000091$ & $1.0231\pm 0.0078$ & $1.0292\pm  0.0011$ &  $1.0342\pm  0.0015$ \\ \midrule
    $\theta_{12}^{q}$ & $0.22704^{+0.00030}_{-0.00029}$ & $0.22704\pm 0.00029$ & $0.22704\pm 0.00029$& $0.22705\pm 0.00029$  \\ \addlinespace
    $\theta_{13}^{q}/10^{-3}$ & $3.71^{+0.13}_{-0.14}$ & $3.79\pm 0.14$ & $3.82\pm 0.14$& $3.85\pm 0.14$\\ \addlinespace
    $\theta_{23}^{q}/10^{-2}$ & $4.181^{+0.047}_{-0.067}$ & $4.270\pm 0.058$ & $4.303\pm 0.058$& $4.337\pm 0.059$ \\ \addlinespace
    $\delta_{\text{CP}}^{q}$ & $1.143^{+0.011}_{-0.011}$ & $1.143\pm 0.011$ & $1.143\pm 0.011$& $1.143\pm 0.011$ \\\bottomrule
  \end{tabular}
\end{table}

\subsection{SM vs.~MSSM Yukawa couplings}

At the SUSY scale (which we set to 1~TeV) two important things happen
at the same time. First of all, loop calculations in the SM
are usually done in the $\MSbar$ scheme which is unsuitable
for supersymmetry. For supersymmetry the $\DRbar$
scheme is preferred.
We will use two-loop RGEs so we have to perform the matching
at the one-loop level. The relevant formulae
for gauge couplings and Yukawa couplings are~\cite{Martin:1993yx}
\begin{align}
  g_{\MSbar} &= g_{\DRbar} \left(1-\frac{g^2}{96 \, \pi^2}\, C(G) \right), \\
  [ Y^i_{\MSbar} ]^{jk} &= [ Y^i_{\DRbar} ]^{jk} \left(1+\frac{g_a^2}{32 \, \pi^2}\, \left( C_a(r_i)+C_a(r_j)-2 \, C_a(r_k)\right) \right),
\end{align}
with $C(G)=N$ and $C(r)=(N^2-1)/2N$ for $SU(N)$. In the MSSM the matching conditions for the Yukawa matrices then read
\begin{align}
  Y^u_{\MSbar} &= Y^u_{\DRbar} \left(1+ \frac{1}{32 \, \pi^2} \left(-\frac{g_1^2}{60}-\frac{3 \, g_2^2}{4}+\frac{8 \, g_3^2}{3} \right) \right), \\
  Y^d_{\MSbar} &= Y^d_{\DRbar} \left(1+ \frac{1}{32 \, \pi^2} \left(-\frac{13 \, g_1^2}{60}-\frac{3 \, g_2^2}{4}+\frac{8 \, g_3^2}{3} \right) \right), \\
  Y^e_{\MSbar} &= Y^e_{\DRbar} \left(1+ \frac{1}{32 \, \pi^2} \left(\frac{9 \, g_1^2}{20}-\frac{3 \, g_2^2}{4}\right) \right).
\end{align}

The second thing is that the Yukawa couplings in the SM and the MSSM
are not the same, not even on tree-level since the MSSM is a two
Higgs doublet model. This introduces a dependence on $\tan \beta$,
the ratio of the two Higgs vevs. Beyond that also finite one-loop corrections
have to be taken into account in the matching. In particular there are
certain pieces which are enhanced by $\tan \beta$ \cite{SUSYthresholds}
and thus can easily make changes of $\mathcal{O}(10 \%)$ or even larger on the Yukawa
couplings.

The approach we will take here is well documented, for instance, in
Refs.~\cite{0804.0717, Spinrath:2010dh, Antusch:2011sq}, 
see also Ref.~\cite{Antusch:2013jca},
so that we will not go into much detail here. Note that we include
the $\tan \beta$ enhanced parts only. This allows a simple
parametrisation of the SUSY threshold corrections in terms of three parameters,
$\epsilon_b$, $\epsilon_q$ and $\epsilon_l$, only
using some rather mild, plausible assumptions on the SUSY breaking
parameters. To be more precise we assume that the squark and slepton mass
matrices are very close to being proportional to the unit matrix
and that the trilinear couplings are hierarchical and
dominated by the 3-3 element. 

For the up-type quarks we can use the tree-level matching relation for SM and MSSM
Yukawa couplings in the $\DRbar$ scheme, 
\begin{equation}
\sin\beta \ Y^u_{\text{MSSM}} = Y^{u}_{\text{SM}} \;,
\label{eq:y_mssm_u}
\end{equation}
since their threshold corrections are proportional to $\cot\beta \ll 1$.

For the Yukawa couplings of the charged leptons and down-type quarks there
are $\tan \beta$ enhanced threshold corrections in the matching formulas
\begin{align}
  \cos\beta (1+\eps_l \tan\beta) \ Y_{\text{MSSM}}^{e}&= Y^{e}_{\text{SM}} \;,\label{eq:y_mssm_e}\\
  \cos\beta \operatorname{diag}\left(1+\eps_q \tan\beta,1+\eps_q \tan\beta,1+\eps_b \tan\beta\right) \ Y_{\text{MSSM}}^{d}&= Y^{d}_{\text{SM}} \;,
  \label{eq:y_mssm_d}
\end{align}
where we have used the conventions as in Ref.~\cite{Antusch:2011sq} but
we defined here $\epsilon_b \equiv \epsilon_A + \epsilon_q$.

Note that $Y_{\text{MSSM}}^{e}$ and $Y_{\text{MSSM}}^{d}$ are not diagonal in our approach
and we determine the corresponding low energy mixing parameters from these matrices after the matching.

In the following, if we do not specify SM or MSSM we refer to MSSM $\DRbar$ quantities.

\subsection{Fitting Procedure}
\label{sec:fittingprocedure}

For every point in the GUT parameter space that is scanned over we extract the fermion observables $\mathcal{O}_i^\text{theo}$ at 1 TeV and compare it to the experimental data $\mathcal{O}_i^\text{exp}$ with the $\chi^2$-function 
\begin{equation}
  \chi^2 = \sum\limits_{i=1}^{15}\left( \frac{ \mathcal{O}_i^\text{theo} -
    \mathcal{O}_i^\text{exp} }{ \sigma_i} \right)^2 + \sum\limits_{i=1}^4
  \chi^2_{\text{PMNS}, i} \;,
  \label{eq:chi2}
\end{equation}
where $\sigma_i$ are the standard deviations of the experimental
observables. 
The first term in Eq.~\eqref{eq:chi2} includes the
masses of quarks (6 observables) and charged leptons (3 observables) as well as the CKM parameters (4 observables) and the mass squared differences of the light neutrinos (2 observables). For these we use a Gaussian error.

The \texttt{NuFit 3.0} results for the mass squared differences of the neutrinos are
\begin{align}
  \Delta m_{21}^2 &= (7.50\pm 0.18)\times 10^{-5} \;\mathrm{eV}^2\,, \\
  \Delta m_{31}^2 &= (2.524\pm 0.0395)\times 10^{-3}\;\mathrm{eV}^2\,.
\end{align}
The other Gaussian observables are listed in Table~\ref{tab:DRbar} in the column '1 TeV'. For reference we also show the corresponding results in the $\MSbar$ scheme in Table~\ref{tab:SM}.

The Gaussian treatment of the errors would be yet inadequate for the PMNS parameters. 
Therefore, for these we use the corresponding $\chi^2$-profiles from
\texttt{NuFit 3.0} \cite{Esteban:2016qun}, which are represented in the second term in Eq.~(\ref{eq:chi2}). 
Our implementation also includes a check for inverted or normal mass 
ordering to ensure the right distribution is used. 
In particular we also include the second
(higher) $\chi^2$ minimum of $\theta_{23}^{l}$. In that way, also the CP phase $\delta_{\text{CP}}^l$
can be included although it is not yet measured directly. 
Note that neutrino observables are treated here as tree-level observables and consequently by
definition do not take part in the
running under the renormalisation group. 
Considering this, we employ the results of \texttt{NuFit 3.0}  
directly at $M_{\text{SUSY}}=1\,\mathrm{TeV}$.

Altogether, we have 19 observables and 22 parameters in the fit, including the three threshold parameters
$\epsilon_b$, $\epsilon_q$ and $\epsilon_l$. As these stem from 1-loop corrections, we allow for them
the following ranges:
\begin{align}
  -0.05 \leq \epsilon_q \leq  0.05 \;, \label{eq:thresholdrange1}\\
  -0.10 \leq \epsilon_b \leq  0.10 \;, \label{eq:thresholdrange2}\\
  -0.03 \leq \epsilon_l \leq  0.03 \;, \label{eq:thresholdrange3} 
\end{align}
see, for example, Refs.~\cite{0804.0717, Antusch:2009gu, Antusch:2011sq, Antusch:2013jca}.
The corrections to quarks are generally larger since they receive SUSY QCD corrections.
Furthermore, $\epsilon_b$ receives another correction from the potentially large stop trilinear
SUSY breaking coupling. One might worry here that the corrections become non-perturbative, since
our scan allows, e.g., $|\epsilon_b \tan \beta| > 1$.
But this is not the case since there are no higher order corrections
$\mathcal{O}( (\epsilon_b \tan \beta)^n )$ with $n\geq2$, see, e.g.,~\cite{Carena:1999py}.

In order to perform the fits we link our \texttt{C++} code to the
Sbplx/Subplex~\cite{NLOPT, Rowan1990} and ISRES algorithms~\cite{NLOPT, Runarsson2005, Runarsson2000}
implemented in the \texttt{NLopt~2.4.2} library~\cite{NLOPT}. 
The extensive numerical fits are performed on the TTP computing cluster. As these are highly time consuming 
we do not include two minor online-updates of the \texttt{NuFit} results which appeared in the meantime.

\section{Results of the SUSY SO(10) Fit to Flavour Data}
\label{sec:Results}

In the following, we present our results for the $\chi^2$ fit of SUSY SO(10) models to the quark
and lepton flavour data given at 1 TeV as described in 
Section~\ref{sec:fittingprocedure}.
First, we present the global minima, which show that threshold corrections are essential to get a good fit in our setup. 
Afterwards, we present the $\chi^2$-profiles for the SUSY threshold parameters.

\subsection{Comparison of the Global Minima} 
\label{sec:globalminima} 

\begin{table}
  \centering
  \caption{\textbf{Results of the global minimisation.} 
    Minimal $\chi^2$, pulls of the observables in the $\DRbar$ scheme
    (for the definition of pulls see Eqs.~(\ref{eq:pull1}) and (\ref{eq:pull2})) for different values of
  $\tan\beta$ as well as our best-fit values for the light neutrino
  masses and SUSY threshold corrections (TC) if included in the fit.}
  \begin{tabular}{lccccccc}\toprule
   & & \multicolumn{3}{c}{w/o TC}  & \multicolumn{3}{c}{with TC} \\
    \cmidrule(lr{.75em}){3-5}  \cmidrule(lr{.75em}){6-8}
  &  $\tan \beta$  & $10$ & $38$ & $50$ & $10$ & $38$ & $50$\\
\cmidrule{2-8}
& $\chi^2$ & 127.0 & 94.69 & 75.43 & 40.37 & 1.74 & 3.74 \\ \midrule
Pulls & $y_u$&	0.19&	0.29&	0.12&	0.01&	0.25&	-0.17\\
& $y_c$&	2.73&	2.71&	2.44&	1.66&	0.19&	0.11\\
& $y_t$&	-2.06&	-2.01&	-1.85&	-1.26&	-0.35&	0.06\\\cmidrule{2-8}
& $y_d$&	-7.42&	-8.08&	-6.30&	-4.53&	0.82&	1.26\\
    & $y_s$&	0.90&	2.12&	1.17&	-0.12&	-0.17&	-0.42\\
& $y_b$&	-0.36&	-0.55&	0.39&	-0.22&	-0.04&	-0.15\\\cmidrule{2-8}
& $\theta_{12}^{q}$&	0.62&	0.45&	0.38&	0.11&	-0.05&	-0.08\\
& $\theta_{13}^{q}$&	-0.71&	-1.03&	-0.42&	2.81&	0.35&	-0.11\\
& $\theta_{23}^{q}$&	-0.38&	-0.58&	0.40&	-1.61&	0.52&	-0.49\\
& $\delta_{\text{CP}}^{q}$&	1.51&	0.64&	0.77&	0.38&	-0.12&	-0.16\\\cmidrule{2-8}
& $y_e$&	0.01&	0.01&	0.04&	0.17&	-0.08&	-0.04\\
& $y_\mu$&	-0.30&	-0.73&	-0.33&	0.28&	-0.15&	0.04\\
& $y_\tau$&	0.42&	0.14&	0.19&	0.30&	-0.17&	0.00\\\cmidrule{2-8}
& $\Delta m_{21}^2$&	0.81&	0.17&	0.49&	-0.30&	-0.05&	0.03\\
& $\Delta m_{32}^2$&	-0.25&	-0.11&	-0.31&	0.19&	0.03&	-0.02\\ \cmidrule{2-8}
& $\theta_{12}^{l}$&	-1.30&	-0.45&	-0.87&	-1.07&	-0.02&	-0.01\\ 
& $\theta_{13}^{l}$&	-4.13&	-0.40&	-0.93&	0.72&	0.00&	-0.04\\
& $\theta_{23}^{l}$&	-5.74&	-2.67&	-4.29&	-0.48&	0.18&	-0.26\\
& $\delta_{\text{CP}}^{l}$&	-1.83&	-1.73&	-1.78&	1.61&	0.56&	1.24\\\midrule
Best-fit values & $m_{\nu,1}$ in meV&	2.4&	2.6&	2.4&	1.8&	2.4&	2.0\\
& $m_{\nu,2}$ in meV&	9.1&	9.1&	9.0&	8.8&	9.0&	8.9\\
&     $m_{\nu,3}$ in meV&	50.2&	50.3&	50.2&	50.3&	50.3&	50.3\\\cmidrule{2-8}
& $\epsilon_q/10^{-2}$&	--&	--&	--&	5.00&	2.80&	4.72\\
& $\epsilon_b/10^{-2}$&	--&	--&	--&	-7.35&	-4.06&	-0.60\\
& $\epsilon_l/10^{-2}$&	--&	--&	--&	-3.00&	-0.60&	0.13\\\bottomrule
  \end{tabular}
  \label{tab:results}.
\end{table}

\begin{figure}
  \centering
  \includegraphics[width=12cm]{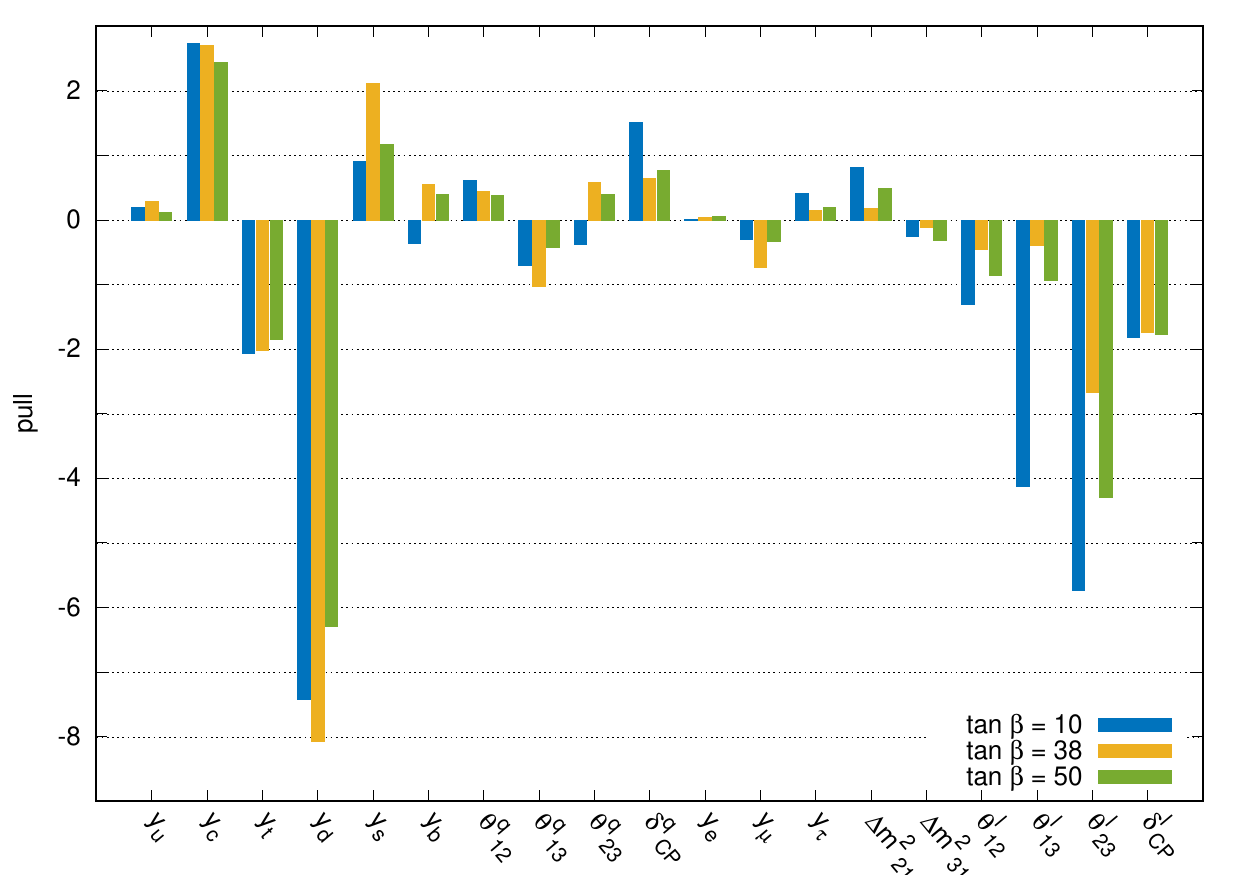}\\
  \includegraphics[width=12cm]{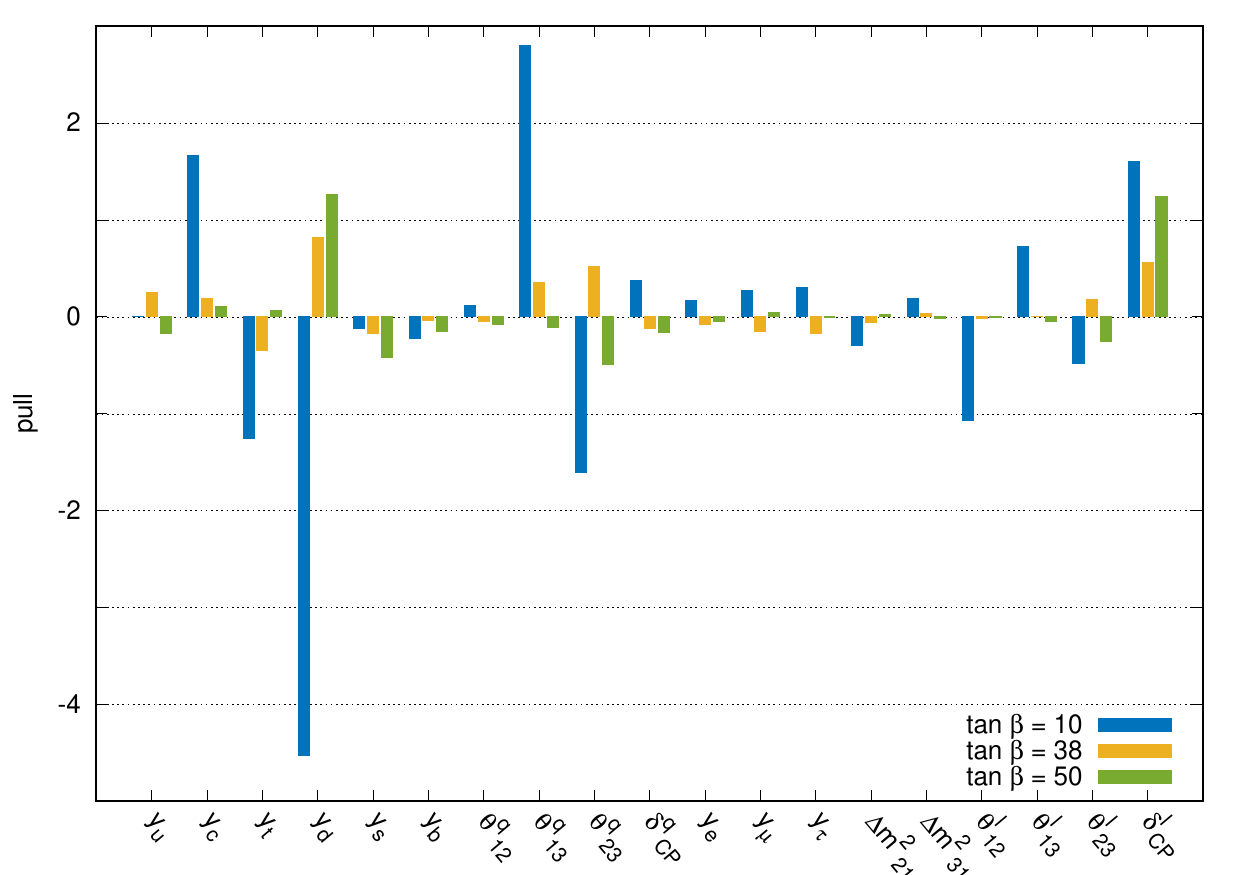}
  \caption{\textbf{Results of the global minimisation.} Pulls of the
    observables for different values of $\tan\beta$ without (top) and
    with (bottom) SUSY threshold corrections. 
    }
  \label{fig:pulls}
\end{figure}

We search for the global minima for three different values of $\tan\beta = 10$, $38$ and $50$ as 
well as both with and without SUSY threshold corrections so that we have six global minima
in total. 
The results of our global minimisation can be found in Table
\ref{tab:results}. 
The table shows the minimal $\chi^2$ from our scan, the pulls of
the SM observables  
as well as the predictions for the light neutrino masses and the needed value of the
threshold corrections in case they are included.
We define the $\mathrm{pull}$ for an observable $\mathcal{O}_i$ with a Gaussian error by
\begin{equation} \label{eq:pull1}
\mathrm{pull}_i = \frac{\mathcal{O}_i^{\mathrm{theo}} - \mathcal{O}_i^{\mathrm{exp}}}{\sigma_i}\,. 
\end{equation}
For the PMNS parameters we use the $\chi^2$-distribution and the best-fit values of \texttt{NuFit} $\mathcal{O}_i^{\mathrm{best-fit}}$ to define
\begin{equation} \label{eq:pull2}
  \mathrm{pull}_i = \operatorname{sign}{(\mathcal{O}_i^{\mathrm{theo}} - \mathcal{O}_i^{\mathrm{best-fit}})}\, \sqrt{ \chi^2_{\mathrm{PMNS,i}}}\,. 
\end{equation}
Additionally, these
pulls are also visualised in the bar charts of Fig.~\ref{fig:pulls}.
The corresponding detailed fit parameters at the GUT scale are given
in Appendix~\ref{sec:gutscaleparameters}.

Without including SUSY threshold corrections we find a poor description of the data by our
SO(10) setup, with best fit points having $\chi^2>70$. It turns out that this is especially triggered
by the high precision of quark mass determinations on the lattice,
in particular, for the down-type quark Yukawa coupling the tension is for all three values of $\tan \beta$ more
than $6\sigma$. There is another major tension in the atmospheric mixing angle where the
modulus of the pull is always larger than
$2$ and even larger than $5$ for $\tan \beta = 10$.

Note that the fit without threshold corrections has as many parameters as observables, and
including threshold corrections 
we have three parameters more than observables so that one might expect to see a perfect fit.  
However, the dependence of the observables on the parameters is highly non-linear, so that the 
expectation of a perfect fit in this case is misleading. 
Indeed, the observables are singular values of complex matrices with a strong
hierarchical structure and are coupled via non-linear differential
equations. This also leads to the fact that not every GUT scale fit parameter
enters the same low-scale observables. For instance, the
1-1 elements of the Yukawa matrices will practically not affect the masses
of the third generation fermions, whereas the 3-3 elements affect
all other couplings and together with the 2-3 elements determine the
mixing between second and third generation.

Having more parameters than observables the statistical interpretation of
$\chi^2$ is not straight forward.
Nevertheless, we can still use the $\chi^2$ information in order
to compare different fit scenarios.
First of all, our results clearly prefer a large $\tan\beta$ which is a well-known result. 
Our minimal SO(10) setup is hardly viable without threshold corrections and for $\tan\beta=10$
even including that corrections,
see Table~\ref{tab:results}.
Therefore, we conclude that sizeable, i.e., percent order SUSY threshold corrections and large
values of $\tan\beta$ are needed in order to be in accordance with data.
Note that for $\tan\beta=10$ at the global minimum we have $\epsilon_q = 0.05$ and
$\epsilon_l = -0.03$ which are both at the 
edges of the respective allowed ranges for these parameters. That means the fit would
prefer larger values of $\eps_q$ and $\epsilon_l$ or larger $\tan \beta$.

Comparing the pulls of the different observables, it is remarkable that besides 
$y_d$ we find large discrepancies in the neutrino observables. This makes
a thorough treatment of the right-handed neutrinos necessary: We treat them in
an effective field theory picture where each of them is integrated out
at its mass scale, as described in detail in Ref.~\cite{Antusch:2005gp}.

Within our setup we always find the neutrinos to have a normal mass hierarchy with
the heaviest neutrino having a mass of about $50\, \mathrm{meV}$.
The sum of light neutrino masses is below the
cosmological bound
$\sum m_\nu \lesssim 0.23 $~eV/$c^2$~\cite{Ade:2015xua} for all global minima.
The same holds obviously for the effective $\beta$-decay mass
\begin{equation}
  m_{\nu,\beta}^2 = \sum\limits_{i=1}^3 \lvert V^{l}_{ei} \rvert^2 m^2_{\nu,i},
\end{equation}
where the current upper bound is $m_{\nu,\beta}<2.05$~eV~\cite{Aseev:2011dq}.
Our best-fit results are also far below the sensitivity of KATRIN \cite{Osipowicz:2001sq} which
is expected to improve this bound by an order of magnitude.

\subsection{Likelihood Profiles for Threshold Corrections}
\label{sec:likelihood}

\begin{figure}
  \centering
  \includegraphics[width=0.49\textwidth]{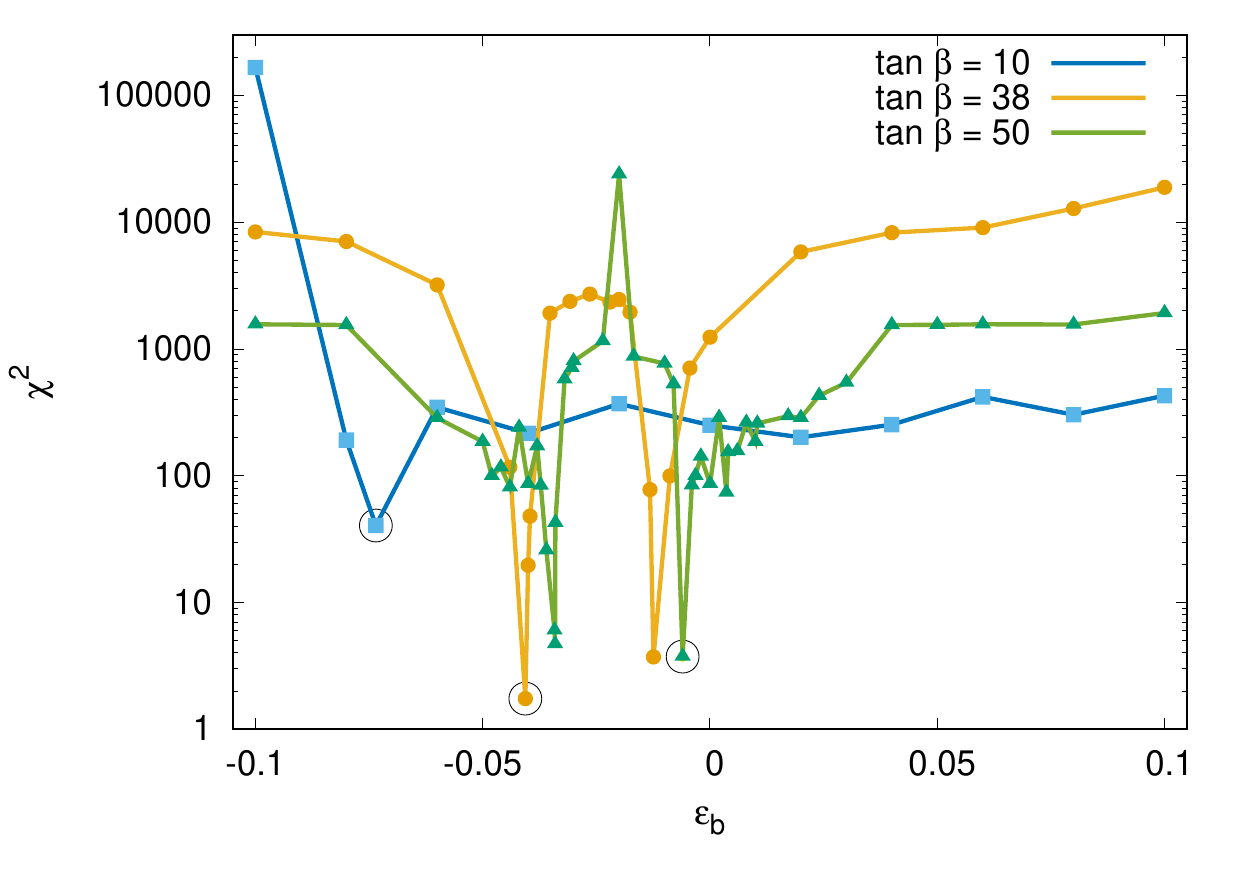}  \hfill
  \includegraphics[width=0.49\textwidth]{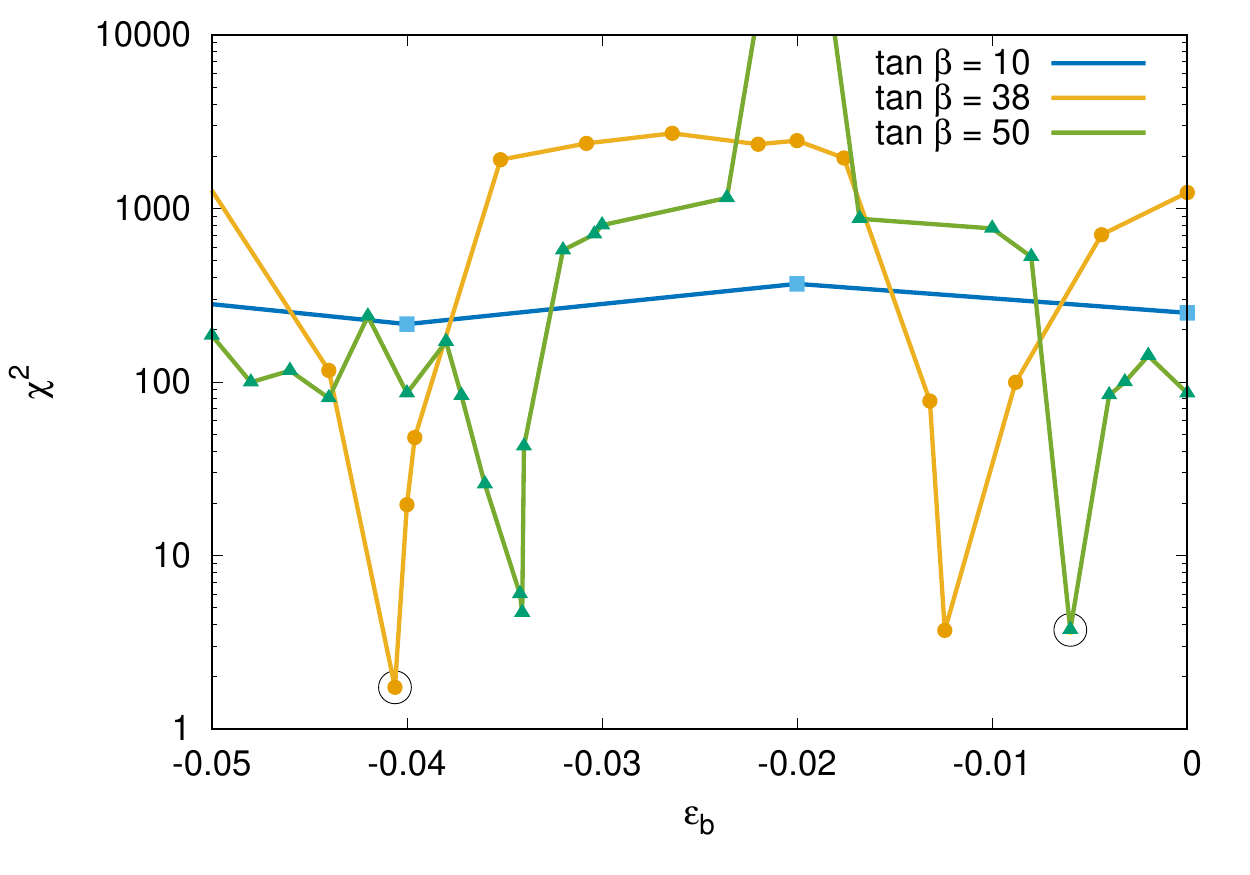}
  \includegraphics[width=0.49\textwidth]{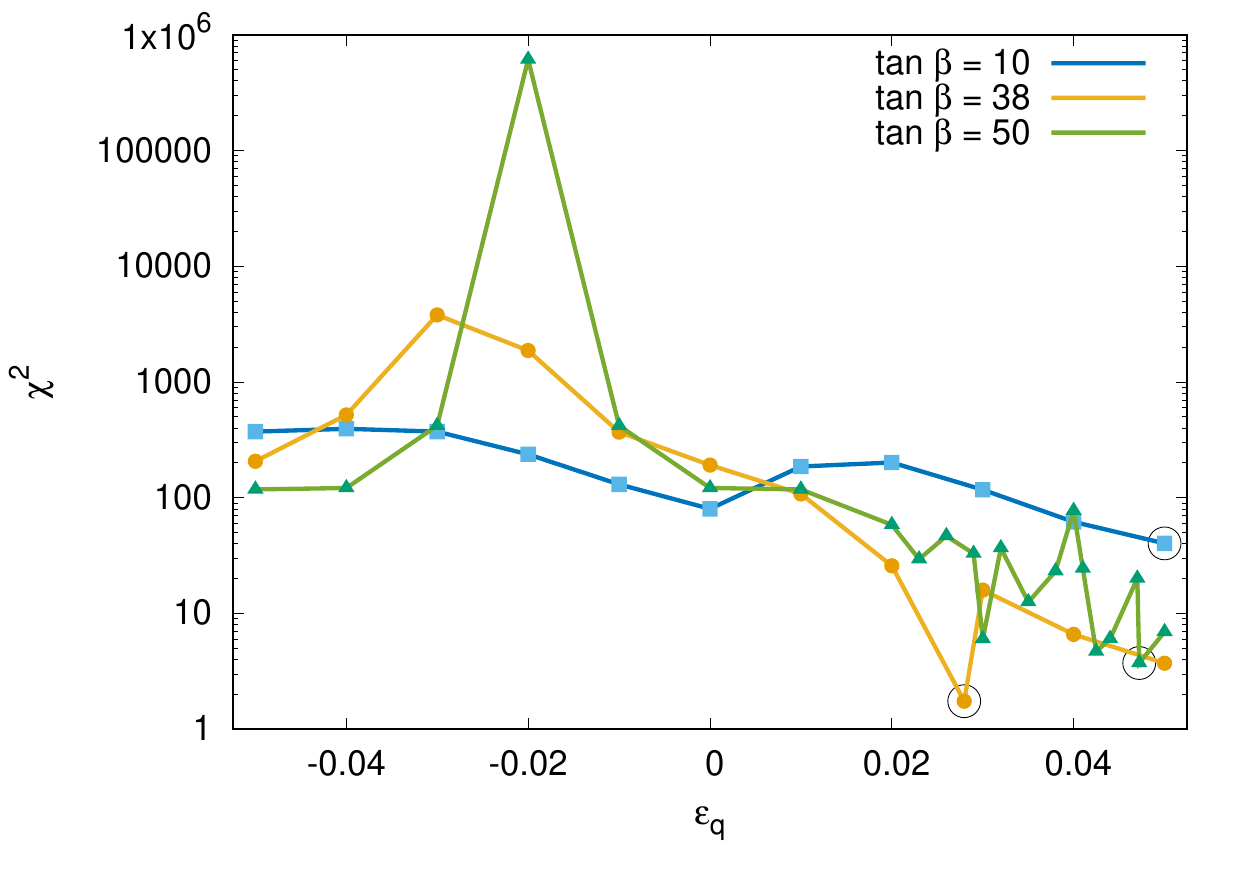}  \hfill
  \includegraphics[width=0.49\textwidth]{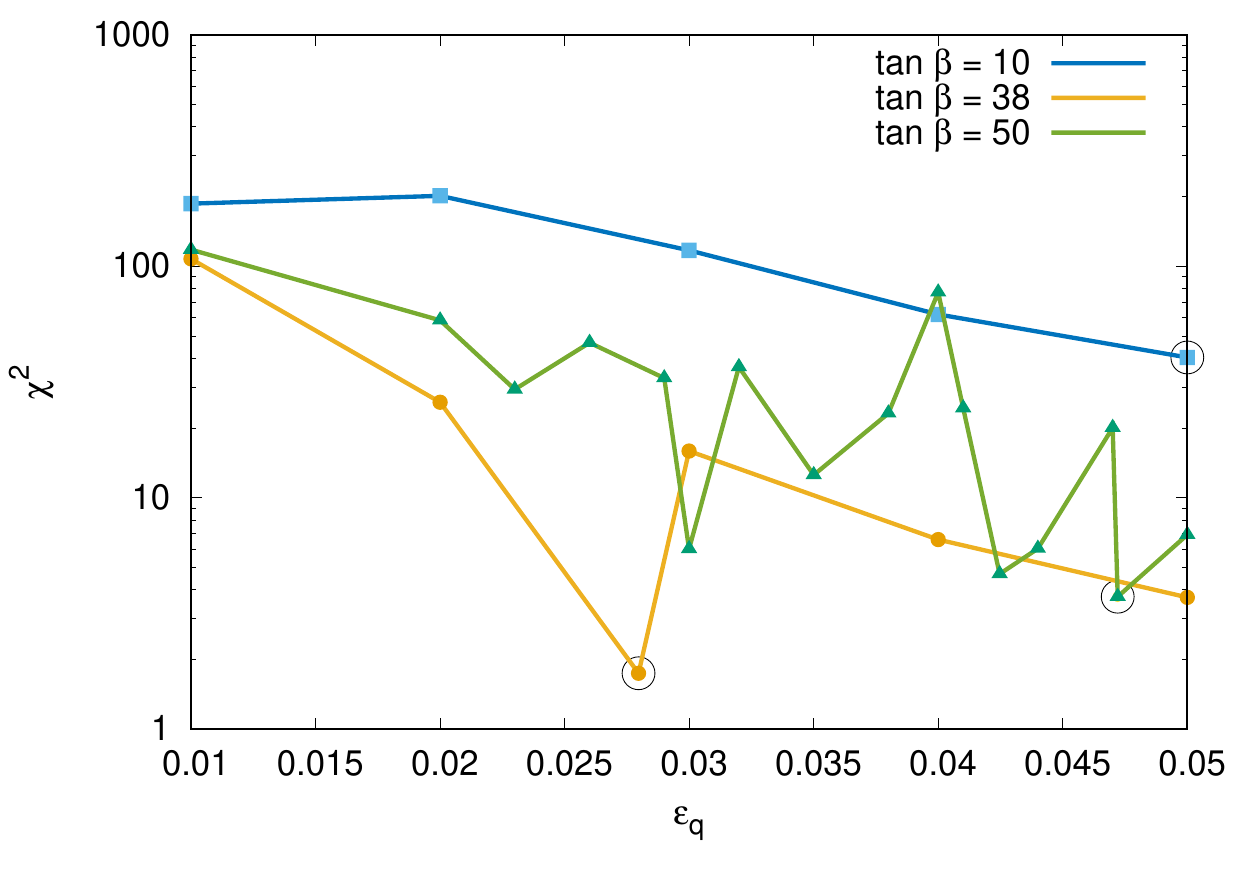} 
  \includegraphics[width=0.49\textwidth]{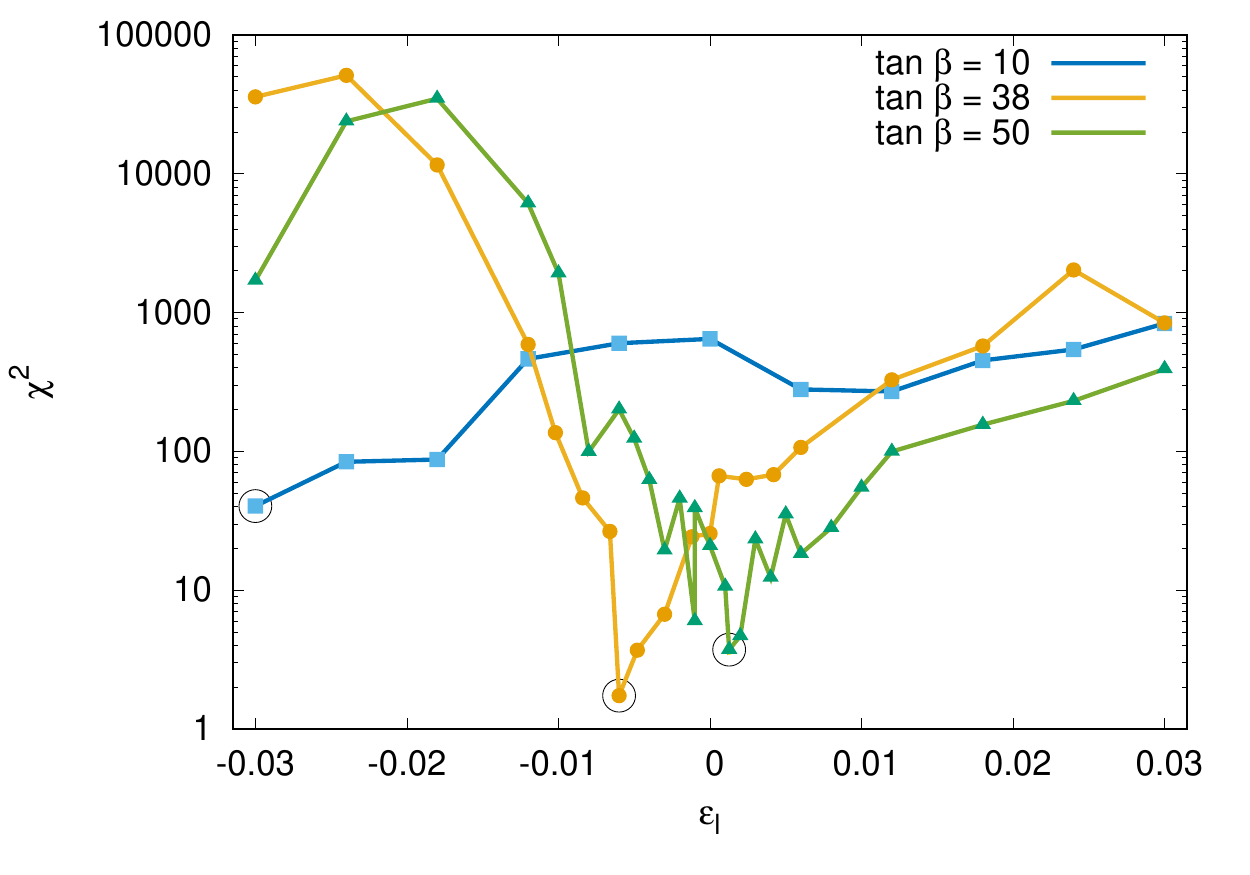} \hfill
  \includegraphics[width=0.49\textwidth]{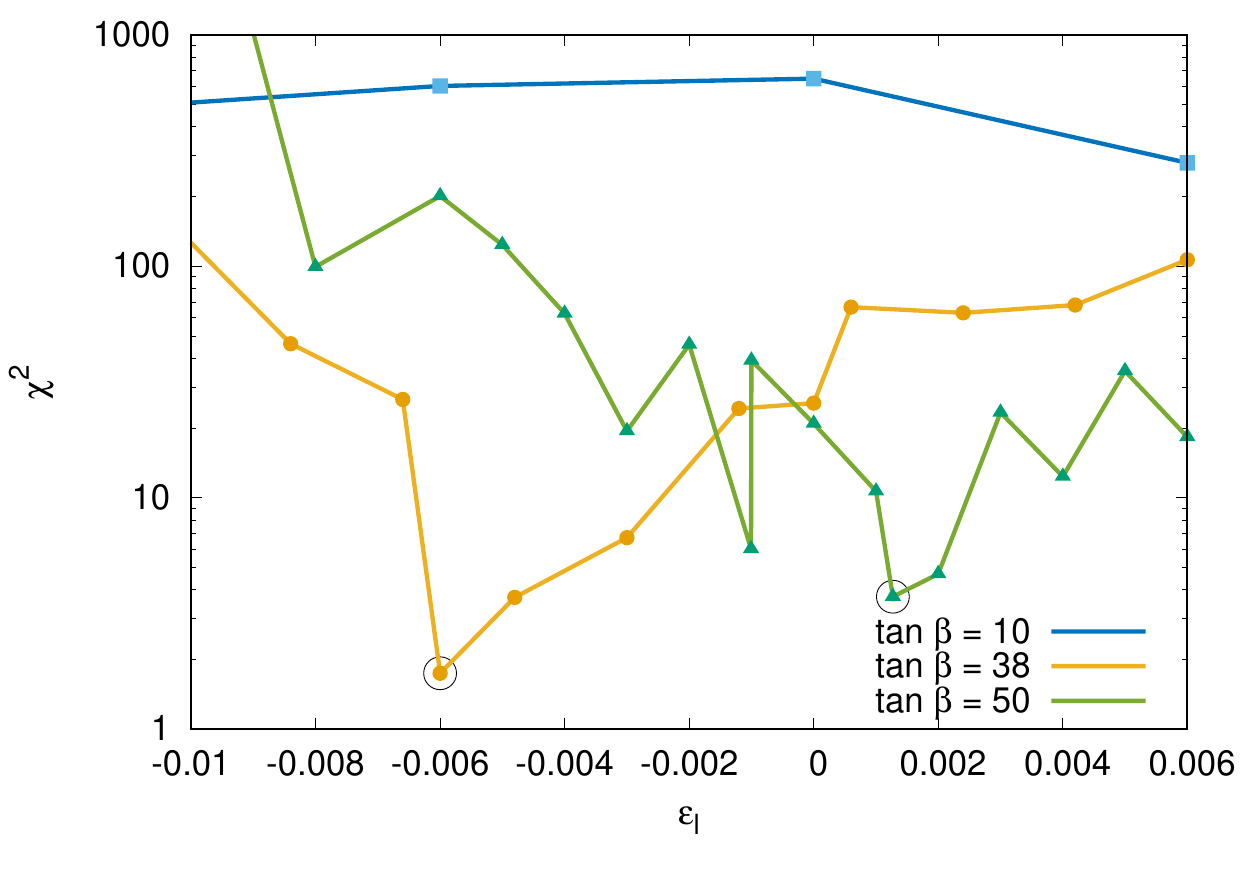}
  \caption{\textbf{Results of the minimisation for fixed values of the threshold corrections.} The circles mark the global minima. The left plots show the full range of the fit whereas the right plots are zoomed around the global minima.
  }
  \label{fig:profiles}
\end{figure}

\begin{figure}
  \centering
  \includegraphics[width=12cm]{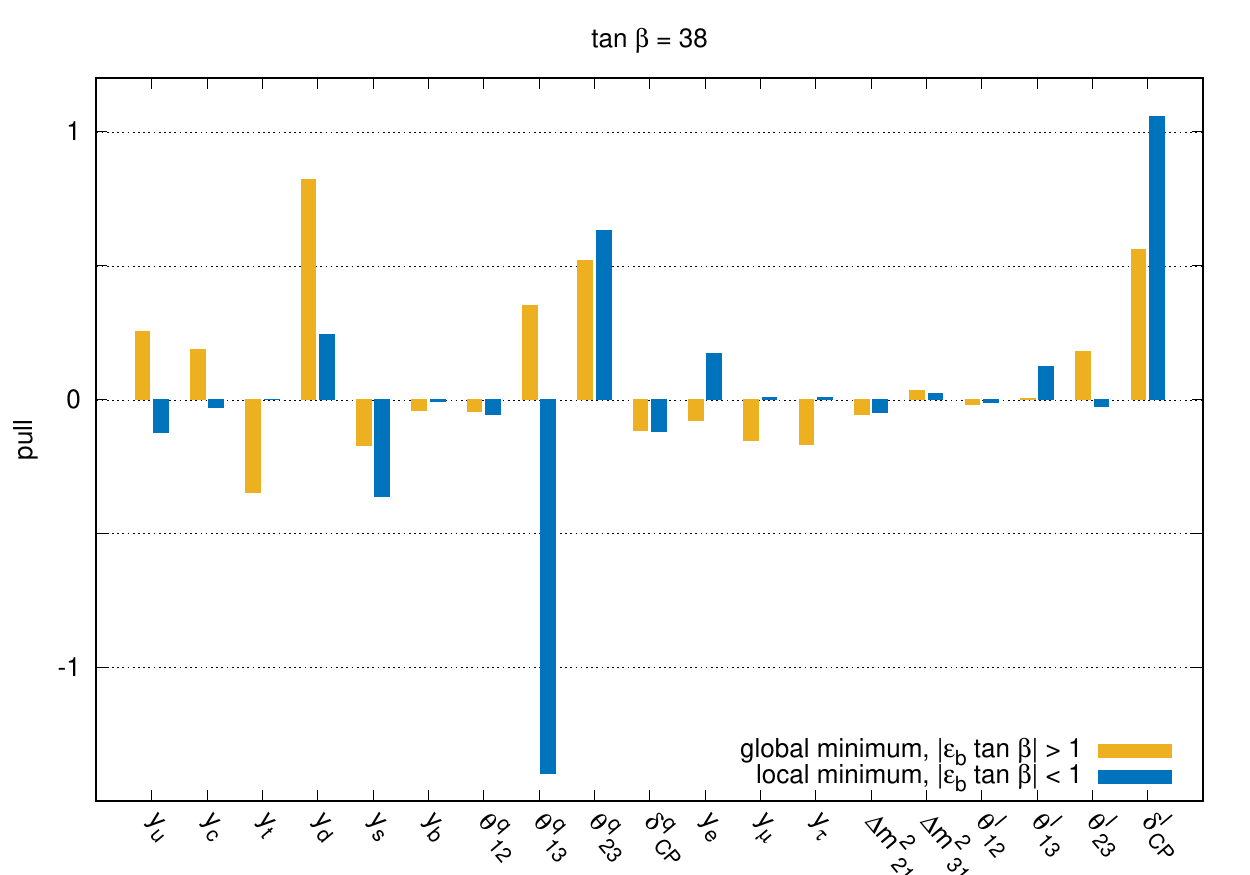}\\
  \includegraphics[width=12cm]{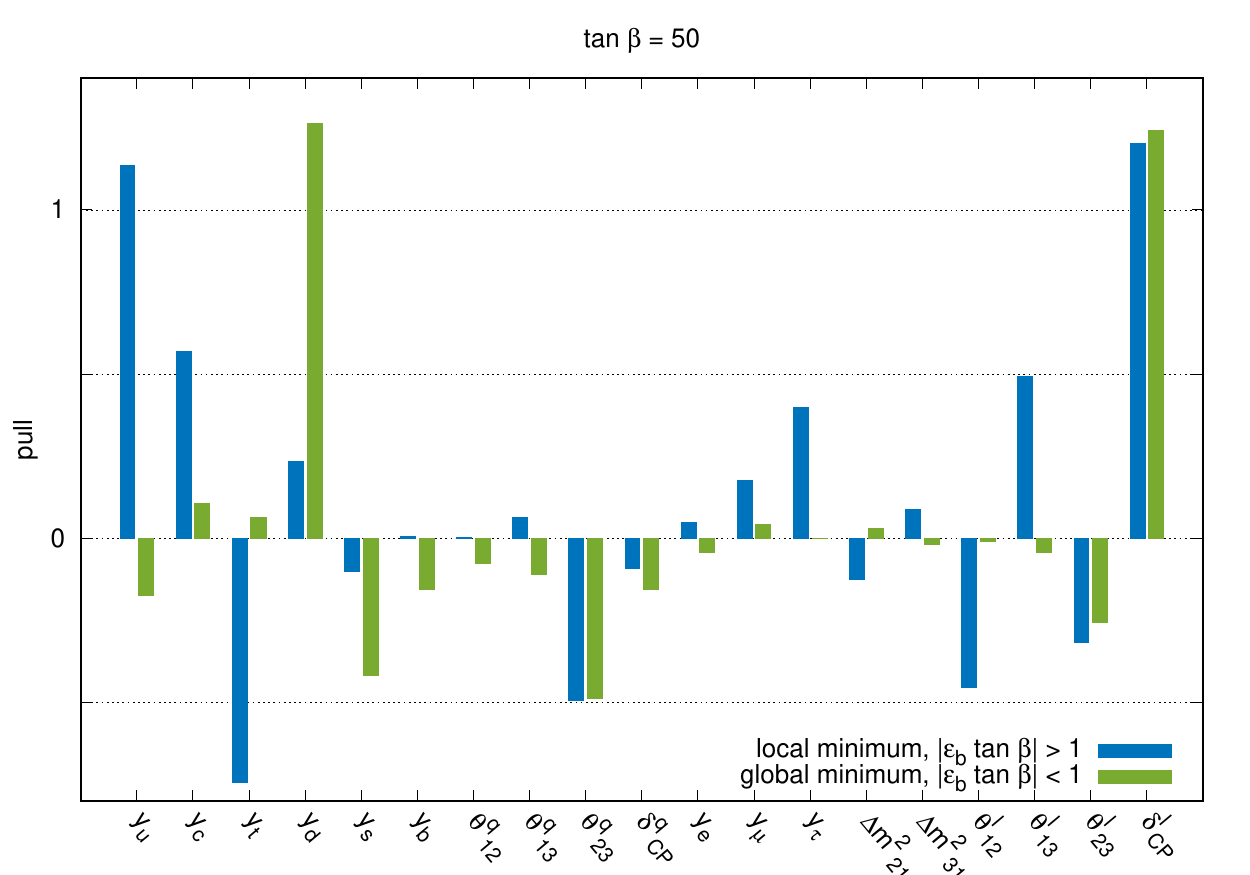}
  \caption{\textbf{Comparison of global minima and local minima.} Shown are the pulls of the observables corresponding to the narrow minima in the $\chi^2$ profiles of Fig.~\ref{fig:profiles}.}
  \label{fig:pulls-local-minima}
\end{figure}

Given the importance of the SUSY threshold corrections for
the goodness of fit of SO(10) models to fermion masses,
see Section~\ref{sec:globalminima}, 
it is an interesting question how sensitive the data is to the magnitude
of the threshold corrections.
This has in particular non-trivial implications for potential
concrete SUSY spectra.
In order to answer this question, we determine the $\chi^2$-profiles of
$\epsilon_b$, $\epsilon_q$ and $\epsilon_l$, which
are shown in Fig.~\ref{fig:profiles}. On the left-hand side we show the
results over the full range that 
we allow, see Eqs.~\eqref{eq:thresholdrange1}--\eqref{eq:thresholdrange3},
while on the right-hand side we zoom into the 
interesting regions around the minima for $\tan\beta=38,\, 50$.
Around the minima we increased the number of points to exclude possible
numerical artefacts and to obtain a better resolution of the profile.

Keeping in mind that the threshold corrections get multiplied by $\tan\beta$, 
see Eqs.~\eqref{eq:y_mssm_e} and \eqref{eq:y_mssm_d}, we first 
like to discuss the case
of $\tan\beta=10$. In this case, the $\chi^2$-profiles have in general less
features than for $\tan\beta=38,\, 50$.
Still, there is a distinct minimum in the $\epsilon_b$-profile and a clear
preference for smaller 
values of $\epsilon_l$ and larger values of $\epsilon_q$. 
Actually, the global minimum lies on the border of the allowed ranges for
$\epsilon_q$ and $\epsilon_l$ as we mentioned before.

Interestingly, for $\tan\beta=38$ and $\tan\beta=50$ we see two
competing narrow minima in the $\chi^2$-profile of $\epsilon_b$.
In fact, our fit fixes the viable values for $\epsilon_b$
to narrow intervals which can be an interesting input 
for SUSY model building and phenomenology. Furthermore, we obtain
that $\epsilon_b<0$ whereas $\epsilon_q>0$.
This means that the fit favours to decrease the bottom Yukawa coupling
and to increase the Yukawa coupling for the strange quark and down quark.

As a curiosity, we note that in our scans the deeper of the two minima
for $\tan\beta=38$ is the one with smaller
$\epsilon_b$ while for $\tan\beta=50$ it is the other way around.
In particular for $\tan \beta = 50$
the minima are almost degenerate with $\chi^2 = 3.7$ and $\chi^2=4.7$.
It might be an interesting observation for SUSY phenomenology, that the two minima competing correspond to either $\epsilon_b \tan \beta < -1$ or $-1 < \epsilon_b \tan \beta < 0$, respectively.

Altogether, it is interesting to note that the largest tensions,
cf.~Figs.~\ref{fig:pulls} and \ref{fig:pulls-local-minima},
are in the down quark mass, up quark mass, $\theta_{13}^q$ and
the leptonic CP phase $\delta^l_{\text{CP}}$. Especially, for the light quark
masses and the leptonic CP phase we expect a significant reduction in the
uncertainties in the next decade which might help to disfavour one of the two
minima or even both minima.
The importance of the leptonic CP phase for SO(10) fits was also noticed
already, e.g., in Ref.~\cite{Babu:2018tfi}.
This again shows that, although we have more
parameters than observables, we can still make quantitative statements.

\section{Summary and Conclusions}
\label{sec:Summary}

SUSY SO(10) is one of the best motivated extensions of the SM and due to its
aesthetics it served as a guiding principle in particle physics for quite a while. Yet,
as no signs of SUSY or proton decay have been seen in experimental searches,
it still remains uncertain if SUSY GUTs are just a dream or if they have indeed anything
to do with nature.

Fermion masses and mixing can give a partial answer to this question.
In this work we have fitted a minimal SO(10) Yukawa sector
to recent data for fermion masses and mixing.
Regarding the available data, compared to most previous
studies there is a
quite significant reduction of uncertainties for light quark masses and neutrino mixing parameters.
These changes are so drastic that minimal SUSY SO(10) without the inclusion of sizeable
SUSY threshold corrections is highly disfavoured nowadays. 

Although it was known for quite a while already that these corrections are generically large and
important \cite{SUSYthresholds} they were usually neglected in fits.
Even in our rather minimal setup the Yukawa sector has already 19 parameters which we fit
to 19 observables. SUSY thresholds plausibly add at least another three parameters which we have
included here increasing the computational needs quite drastically.

In fact, we did not only include them but also provide for them $\chi^2$-profiles
for the first time to our knowledge. These turn out to be particularly interesting for
the third generation
parameter $\epsilon_b$. We find two rather narrow minima for $\tan \beta = 38$ and $50$ each, all of 
them with an acceptable $\chi^2 < 5$.
These precise predictions for $\epsilon_b$ have consequences
for the SUSY spectrum and phenomenology which nevertheless goes beyond the scope of the current work.
Unfortunately, we could not identify here a general set of observables which allows us to distinguish
between the two solutions for both $\tan \beta$ values.
In principle, once the SUSY spectrum is known we can directly
calculate the $\epsilon$-parameters which would provide a definite answer.

Until then a more precise determination of the light quark masses and neutrino mixing
observables will help to challenge some of the minima which we have found here.
The good thing is that we expect a significant progress for these observables in the
next years before a detailed knowledge of the SUSY spectrum could realistically be available.

In particular, we would like to highlight here the
role of the leptonic Dirac CP violating phase $\delta_{\text{CP}}^l$ which creates 
a considerable contribution to the $\chi^2$ for most of the found minima. Hence, with the measurement 
of low energy leptonic CP violation we can test SUSY SO(10).

In summary, we see that with the emergence of precision neutrino data and improved lattice
calculations SUSY SO(10) can be challenged. So that in the near future we might find out, if we
have been foolishly dreaming or following a path towards a deeper understanding of nature.

\section*{Acknowledgements}

TD acknowledges support by the doctoral school GRK 1694 (DFG).
SS is supported by a DFG Forschungsstipendium under contract no.\ SCHA 2125/1-1.
MS is supported by the Ministry of Science and Technology (MOST) of Taiwan 
under grant number MOST 107-2112-M-007-031-MY3.

\appendix

\section{SM Observables at Various Scales}
\label{sec:observables}

The quark masses (except for the top quark) and the value of $\alpha_s^{(5)}$ are taken from Ref.~\cite{Aoki:2016frl}, as described in Section \ref{sec:fermionmasses}. For the mass of the top quark we use the PDG average $m_{t,\text{pole}}=174.2 \pm 1.4\, \text{GeV}$ \cite{Olive:2016xmw}. At two-loop accuracy also the quartic Higgs coupling contributes to the RGE running of the Yukawas in the SM
\begin{equation}
  \lambda = \frac{2 \, m_{\text{Higgs}}}{ v^2} \,.
\end{equation}
This convention is in accordance with the convention used in the two-loop RGEs, i.e.\ 
the quartic term in the SM Higgs potential is given by $\lambda (\phi^\dagger\phi)^2/4$.
We calculate $\lambda$ from the Higgs mass measurement $m_{\text{Higgs}} = 125.09 \pm 0.24\; \text{GeV}$ \cite{Olive:2016xmw}.

The values for the charged lepton Yukawa couplings and the electroweak gauge couplings have already been given in Ref.~\cite{Antusch:2013jca}
\begin{align}
  10^6 \times Y_e(M_Z) &= 2.794745 \pm 0.0000155 \;,\\
  10^4 \times Y_\mu(M_Z) &= 5.899863 \pm 0.0000185 \;,\\
  10^2 \times Y_\tau(M_Z) & = 1.002950 \pm 0.0000905 \;,\\
  g_1(M_Z) &= 0.461425 \pm 0.0000435 \;,\\
  g_2(M_Z) &= 0.65184 \pm 0.000175 \;.
\end{align}
For the CKM matrix  we use the updated results of the CKMfitter group as presented at ICHEP 2016 \cite{Charles:2004jd}. We translate from the Wolfenstein parametrisation to the parametrisation of the PDG and
obtain
\begin{align}
  \theta_{12}^q &= 0.227035 \pm 0.000293\\
  \theta_{13}^q &= 0.003712 \pm 0.000139\\
  \theta_{23}^q &= 0.0418112 \pm 0.000567\\
  \delta_{\text{CP}}^q &= 1.1430 \pm 0.0108
\end{align}
In Table \ref{tab:SM} all these values are given at 1 TeV, 3 TeV and 10 TeV, respectively. The procedure of RGE running is described in Sec.~\ref{sec:fermionmasses}. For use within SUSY models the values have also been converted to the $\DRbar$ scheme, cf. Table \ref{tab:DRbar}.

\section{GUT Scale Parameters for the Global Minima}
\label{sec:gutscaleparameters}

For reference, we give here the coordinates of our global minima to a high numerical
precision, which allows the
reproduction of the results given in Table~\ref{tab:results}.
We determine the GUT scale by minimising 
the sum of the squared differences of the gauge couplings. The central
values of the Yukawa couplings are included in the running to
the GUT scale. At $M_{\text{GUT}} = 1.35276\times 10^{16}$ we find
\begin{equation*}
   g_1 = 0.706132, \qquad  g_2 = 0.708008, \qquad  g_3 = 0.705465.
\end{equation*}
In all fits, these values remain fixed. The parametrisation of the
MSSM Yukawa matrices is given in
Eqs.~\eqref{eq:Yu_GUT}-\eqref{eq:Ka_GUT}. If SUSY threshold corrections are
included, their value at $M_{\text{SUSY}}$ is also given below. The parametrisation of the
SO(10) Yukawa couplings is
\begin{equation*}
  \label{eq:1}
  Y_{10} = \operatorname{diag}\left( H_1, H_2, H_3 \right), \qquad
  Y_{126} = \begin{pmatrix} F_1 & F_2 & F_3 \\ F_2 & F_4 & F_5 \\ F_3 & F_5 & F_6 \end{pmatrix}.
\end{equation*}

\subsection{Global Minima without SUSY Threshold Corrections \label{sec:globalwocorr}}

We list here our fit results for the coordinates of the global minima without SUSY threshold corrections for 
different values of $\tan\beta$. For reference, we use a high numerical precision.

\paragraph{$\tan\beta=10$, no threshold corrections, $\chi^2 = 127.016$}
\begin{align*}
  r &= 8.93109\;, 
& s &= 0.40219 + 0.0537018 \,\ii\;,\\
r_R &= 5.7222\times 10^{14}\,\mathrm{GeV}\;,
&H_1 &= -8.13678\times 10^{-6}\;, \\ H_2 &= 0.000546749\;, & H_3 &= 0.0587967\;,\\
F_1 &= (3.62943 + 1.11577 \,\ii)\times 10^{-5}\;, & F_2 &= (-4.86448 -6.8181 \,\ii)\times 10^{-5}\;,\\
F_3 &= -8.669\times 10^{-5} + 0.00101114\,\ii\;, & F_4 &= -0.00122229 + 0.000483037\,\ii\;,\\
F_5 &= 0.00229999 -0.00243838\,\ii\;, & F_6 &= -0.00436435 + 0.000144527\,\ii\;.
\end{align*}

\paragraph{$\tan\beta=38$, no threshold corrections, $\chi^2 = 94.6859$}
\begin{align*}
r &= 2.09333\;, 
& s &= 0.361579 + 0.00221939\,\ii\;,\\
r_R &= -1.01454\times 10^{14}\,\mathrm{GeV}\;,
&H_1 &= -3.76306\times 10^{-5}\;, \\ H_2 &= 0.0017972\;, & H_3 &= 0.254405\;,\\
F_1 &= 0.000100472 + 0.000137311\,\ii\;, & F_2 &= 3.1103\times 10^{-5} + 0.000457557\,\ii\;,\\
F_3 &= -0.00260288 + 0.0039351\,\ii\;, & F_4 &= -0.00474944 + 0.00161864\,\ii\;,\\
F_5 &= -0.0107501 + 0.00940057\,\ii\;, & F_6 &= -0.0119578 -0.047225\,\ii\;.
\end{align*}  

\paragraph{$\tan\beta=50$, no threshold corrections, $\chi^2=75.428$}
\begin{align*}
r &= -1.351\;, 
& s &= 0.380723 + 0.0112833\,\ii\;,\\
r_R &= 7.08356\times 10^{13}\,\mathrm{GeV}\;,
&H_1 &= -5.44756\times 10^{-5}\;, \\ H_2 &= 0.00295336\;, & H_3 &= 0.418203\;,\\
F_1 &= 0.000198815 + 0.000167965\,\ii\;, & F_2 &= 0.000197095 + 0.00060758\,\ii\;,\\
F_3 &= 0.00302059 -0.00664685\,\ii\;, & F_4 &= -0.00713153 + 0.00251093\,\ii\;,\\
F_5 &= 0.0167445 -0.0153386\,\ii\;, & F_6 &= -0.0253209 -0.0530162\,\ii\;.
\end{align*}

\subsection{Global Minima with SUSY Threshold Corrections}

Here we give our fit results for the coordinates of the global minima including SUSY threshold corrections, 
again with a high numerical precision, see above.

\paragraph{$\tan\beta=10$, with threshold corrections, $\chi^2 = 40.3671$}
\begin{align*}
r &= 2.68759\;, 
& s &= 0.146006 -0.258778\,\ii\;, \\
r_R &= 4.05262\times 10^{13}\,\mathrm{GeV}\;,
& H_1 &= 5.28137\times 10^{-6}\;, \\ H_2 &= 0.000694187\;, & H_3 &= 0.186542\;,\\
F_1 &= (-7.52986 -11.2321\,\ii)\times 10^{-6}\;, & F_2 &= (-5.92321 -9.76299\,\ii)\times 10^{-5}\;,\\
F_3 &= 0.000578945 + 0.000363829\,\ii\;, & F_4 &= 6.89316\times 10^{-5} -0.000631902\,\ii\;,\\
F_5 &= 0.0100828 + 0.000256401\,\ii\;, & F_6 &= 0.0424372 + 0.028391\,\ii\;,\\
  \epsilon_q &=  0.05000\;, &\epsilon_b &= -0.07348963\;, \\\epsilon_l &= -0.02999953\;. &&
\end{align*}

\paragraph{$\tan\beta=38$, with threshold corrections, $\chi^2 = 1.74389$}
\begin{align*}
  r &= 0.739487\;, 
& s &= 0.171775 -0.0527206\,\ii\;, \\
r_R &= 3.00163\times 10^{12}\,\mathrm{GeV}\;, 
& H_1 &= 1.74192\times 10^{-5}\;, \\ H_2 &= 0.00263311\;, & H_3 &= 0.873783\;,\\
F_1 &= (-4.52508 -1.40042)\times 10^{-5}\,\ii\;, & F_2 &= -0.000670712 -0.000352788\,\ii\;,\\
F_3 &= 0.00257905 + 0.000499293\,\ii\;, & F_4 &= -0.00325008 -0.0013532\,\ii\;,\\
F_5 &= 0.0128581 + 0.0601288\,\ii\;, & F_6 &= 0.474974 + 0.216717\,\ii\;,\\
  \epsilon_q &= 0.02796541\;, &\epsilon_b &= -0.04061278\;, \\\epsilon_l &= -0.006000\;. &&
\end{align*}

\paragraph{$\tan\beta=50$, with threshold corrections, $\chi^2 = 3.73552$}
\begin{align*}
r &= 0.699764\;, 
& s &= 0.137602 -0.0279775\,\ii\;, \\
r_R &= 2.33\times 10^{12}\,\mathrm{GeV}\;,
& H_1 &= 1.21069\times 10^{-5}\;, \\ H_2 &= 0.00246264\;, & H_3 &= 0.988718\;,\\
F_1 &= (-4.09545 -1.24086\,\ii)\times 10^{-5}\;, & F_2 &= 0.000472532 + 0.00031884\,\ii\;,\\
F_3 &= 0.00483506 + 0.00119957\,\ii\;, & F_4 &= -0.00190458 + 0.00028723\,\ii\;,\\
F_5 &= -0.0316564 -0.0584933\,\ii\;, & F_6 &= 0.607543 + 0.0633526\,\ii\;,\\
\epsilon_q &= 0.04720754\;, &\epsilon_b &= -0.006000\;, \\\epsilon_l &=  0.001271233\;. &&
\end{align*}

\subsection{Local Minima with SUSY Threshold Corrections}

Here we give our fit results for the coordinates of the local minima of $\epsilon_b$
including SUSY threshold corrections, again with a high numerical precision, see above.

\paragraph{$\tan\beta=38$, with threshold corrections, $\chi^2 = 3.71108$}
\begin{align*}
  r &= 0.670043\;, & s &= 0.11989 - 0.0413042\,\ii\;,\\
  r_R &= 2.03128 \times 10^{12}\,\mathrm{GeV}\;, &H_1 &= 1.09545\times 10^{-5}\;, \\
  H_2 &= 0.00255376\;, & H_3 &= 1.05489\;,\\
  F_1 &= -3.38058\times 10^{-5} -5.62988\times 10^{-10}\,\ii\;, & F_2 &= -0.00040558 - 0.000269377\,\ii\;,\\
  F_3 &= 0.00523839 + 0.000883983\,\ii\;, & F_4 &= -0.00226047 + 0.000662944\,\ii\;,\\
  F_5 &= 0.0259931 + 0.063712\,\ii\;, & F_6 &= 0.622678 + 0.0958844\,\ii\;,\\
  \epsilon_q &= 0.04999992\;, &\epsilon_b &= -0.01243252\;, \\\epsilon_l &= -0.0048000\;. &&
\end{align*}

\paragraph{$\tan\beta=50$, with threshold corrections, $\chi^2 = 4.6864$}
\begin{align*}
  r &= 0.719454\;,  & s &= 0.17033 -0.0410235\,\ii\;, \\
  r_R &= 2.89057\times 10^{12}\,\mathrm{GeV}\;, &H_1 &= 1.14972\times 10^{-5}\;,\\
  H_2 &= 0.00276048\;, & H_3 &= 0.898288\;,\\
  F_1 &= (-3.87624 -1.31146\,\ii)\times 10^{-5}\;, & F_2 &= -0.000600001 -0.000317934\,\ii\;\\
  F_3 &= -0.00275464 -0.00102727\,\ii\;, & F_4 &= -0.00375989 -0.00115394\,\ii\;\\
  F_5 &= -0.0164547 -0.0618966\,\ii\;, & F_6 &= 0.552 + 0.0273906\,\ii\;\\
  \epsilon_q &= 0.4246740\;, &\epsilon_b &= -0.3408847\;, \\\epsilon_l &= 0.0020000\;. &&
\end{align*}

\end{document}